\documentclass[aps,prl,twocolumn,showpacs,superscriptaddress]{revtex4}
\usepackage{graphicx}
\usepackage{subfigure}
\usepackage{amssymb}

\newcommand {\pT}{p_{\rm T}}
\newcommand {\dphi}{\Delta\phi}

\begin{document}
\title{A new correlation method to identify and separate charm and bottom production processes at RHIC}

\author{Andr\'e Mischke\footnote{Contact email: a.mischke@uu.nl}}

\address{Institute for Subatomic Physics, Faculty of Science, Utrecht University, Princetonplein 5, 3584 CC Utrecht, the Netherlands.}

\date{\today}

\begin{abstract}
Electrons from semileptonic decays of heavy-flavor mesons ($D$ and $B$) allow to study the energy loss of heavy-quarks in nuclear collisions at $\sqrt{s}$~=~200 GeV at RHIC. Since pQCD calculations have shown that the crossing point where bottom decay electrons start to dominate over charm decay electrons is largely unknown, an urgent need arises to access the relative contributions independently. A correlation method is proposed to identify and separate charm and bottom production processes on a statistical basis through tagging of their decay electrons and open charmed mesons. The feasibility for this method is demonstrated using PYTHIA and MC@NLO simulations.
The latter allows to estimate the complete NLO contributions, including e.g. gluon-splitting diagrams.
\end{abstract}

\pacs{23.70.+j, 24.10.Lx, 25.75.Cj, 25.75.Gz}


\maketitle

\section{Introduction}
Energy loss of partons is predicted to be a sensitive probe of the matter created in high energy nuclear collisions since its magnitude depends strongly on the color charge density of the matter traversed. 
In particular, the understanding of the flavor dependent coupling of quarks and their fragmentation functions provides key tests of parton energy-loss models and, thus, yields profound insight into the properties of the produced highly-dense strongly interacting matter.
Measurements at the Relativistic Heavy-Ion Collider (RHIC) at Brookhaven National Laboratory have revealed large medium-induced suppression at high transverse momentum (high $\pT$) of both the inclusive hadron yields and of back-to-back hadron pairs~\cite{whitepapers}.
The principal energy loss mechanism underlying these effects is commonly thought to be medium-induced gluon Bremsstrahlung, which is expected to dominate collisional (elastic) energy loss for very energetic partons~\cite{Theo:Wang}.

Due to their large mass ($m >$ 1 GeV/c$^2$), heavy quarks (charm and bottom) are believed to be primarily produced by hard scattering processes (high momentum transfer) in the early stage of the collision and, therefore, are sensitive to the initial gluon density~\cite{Theo:Lin}.
Heavy-quark production by initial state gluon fusion also dominates in nuclear collisions where many, in part overlapping nucleon-nucleon collisions occur~\cite{Star:charmpaper}. 
Heavy-quark production by thermal processes later in the collision is low since the expected energy available for particle production in the medium ($\sim$0.5 GeV) is smaller than the energy needed to produce a heavy-quark pair ($>$ 2.4 GeV). 
Theoretical models based on perturbative Quantum Chromodynamics (pQCD) predicted that heavy quarks should experience a smaller amount of radiative energy loss in the medium than light quarks when propagating through the extremely dense medium due to the suppression of small angle gluon radiation~\cite{Theo:deadcone,Theo:magd}.

The energy loss of heavy-quark mesons is currently studied through the measurements of the $\pT$ spectra of their decay electrons. 
At high $\pT$, this mechanism of electron production is dominant enough to reliably subtract other sources of electrons like conversions from photons and $\pi^0$ Dalitz decays.
RHIC measurements in central Au+Au collisions have shown that the high $\pT$ yield of electrons from semileptonic charm and bottom decays is suppressed relative to properly scaled proton-proton collisions, usually quantified in the nuclear modi\-fication factor ($R_{AA}$)\cite{Star:npe, Phe:npe}. 
This factor exhibits an unexpectedly similar amount of suppression as observed for light-quark hadrons, suggesting substantial energy loss of heavy quarks in the produced medium.
Energy-loss models incorporating contributions from charm and bottom do not explain the observed suppression sufficiently~\cite{Theo:Magda, Theo:Nesto}. 
Although it has been realized that energy loss by elastic parton scattering causing collisional energy loss is probably of comparable importance to energy loss by gluon radiation~\cite{Theo:Wicks, Theo:Hess}, the quantitative description of the suppression is still not satisfying.
Furthermore, it has been shown that collisional dissociation of heavy mesons in the medium may be significant in heavy-ion collisions~\cite{Theo:Adil}.
However theoretical models which include energy loss from charm only describe the observed suppression reasonably well~\cite{Theo:Nesto}. 

The observed discrepancy between data and model calculations could indicate that the $B$ dominance over $D$ mesons starts at higher $\pT$ as expected.
Theoretical calculations implying pQCD have shown that the crossing point where bottom decay electrons starts to dominate over charm decay electrons is largely unknown~\cite{Theo:Matteo,Vogt08}. 
Therefore, the relative contributions from charm and bottom meson decays to electrons have to be determined separately.

This paper reports a new correlation method using azimuthal angular correlations of heavy-quark decay electrons and open charmed mesons, which yields important information about the underlying production mechanism. 

\begin{figure}[t]
\begin{center}
\subfigure{\includegraphics[width=0.261\textwidth]{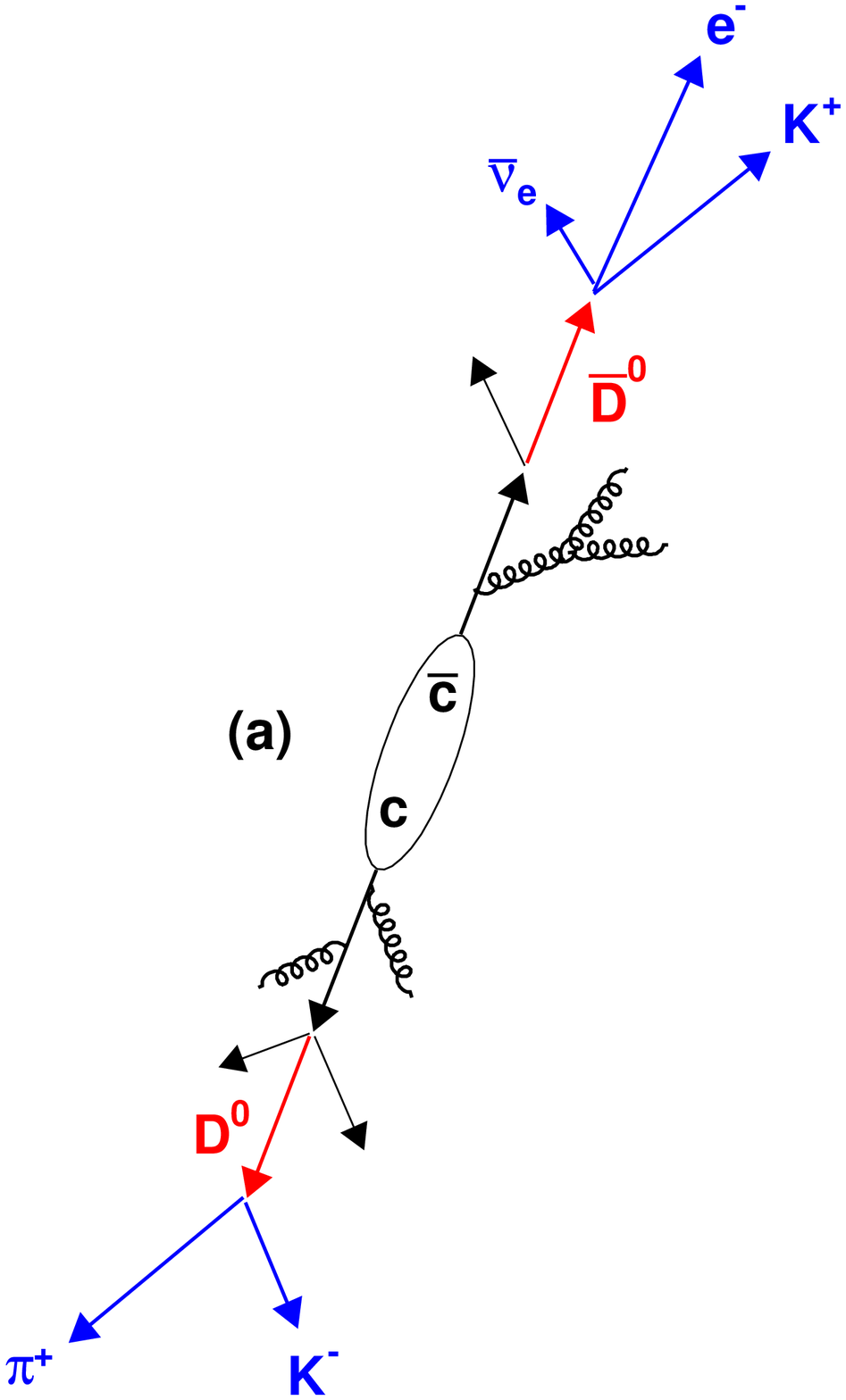}}
\hspace{-9.3mm}
\subfigure{\includegraphics[width=0.261\textwidth]{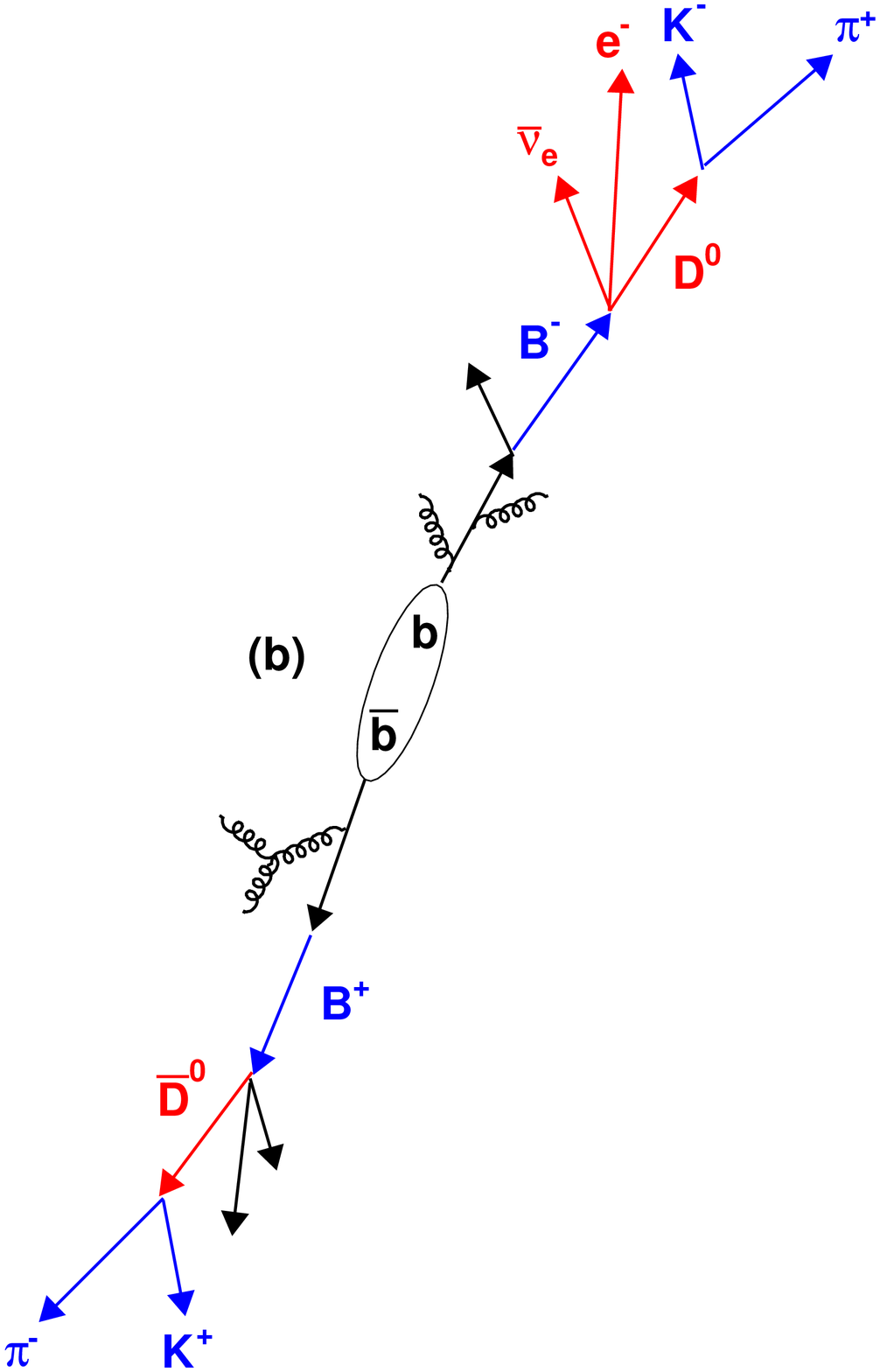}}
\caption{
(Color online) Schematic view of the fragmentation of (a) a $c{\bar c}$ and (b) a $b{\bar b}$ pair.}
\label{fig:11}
\end{center}
\end{figure}

\section{Correlation method}
In Quantum-Chromodynamics, flavor conservation implies that heavy quarks are produced in quark anti-quark pairs ($c{\bar c}$ and $b{\bar b}$). A more detailed understanding of the underlying production process may be obtained from events in which both heavy-quark particles are detected. 
Due to momentum conservation, these heavy-quark pairs are correlated in relative azimuth ($\dphi$) in the plane perpendicular to the colliding beams, leading to the characteristic back-to-back oriented sprays of particles (dijet). A dijet signal appears in the azimuthal correlation distribution as two distinct back-to-back Gaussian-like peaks around $\dphi = 0$ (near-side) and $\dphi = \pi$ (away-side).
The correlation in their azimuthal opening angle survives the fragmentation process to a large extent in $p+p$ collisions. Angular correlations of pairs of high $\pT$ particles have successfully been used to study on a statistical basis the properties of the produced jets~\cite{whitepapers}.

In this correlation method, charm and bottom production events are identified using the characteristic decay topology of their jets. 
Charm quarks predominantly hadronize directly to $D^0$ mesons ($c\rightarrow D^0+X$, BR = $56.5\pm3.2\%$) while bottom quarks produce $D^0$ via an intermediate $B$ meson 
($b\rightarrow B^{-} / \overline{B^0} / \overline{B_{s}^{0}} \rightarrow D^0+X$, BR = $59.6\pm2.9\%$)~\cite{PDG}. 
The branching ratio for charm and bottom quark decays into electrons is 9.6$\%$ and 10.86$\%$, respectively.
While triggering on the so-called leading electron (trigger side), the balancing heavy quark, identified by the $D^0$ meson ($D^0\rightarrow K^{-}\pi^{+}$, BR = 3.89$\%$), is used to determine the underlying production mechanism (probe side). 

\begin{figure}[t]
\begin{center}
\subfigure{\includegraphics[width=0.23\textwidth]{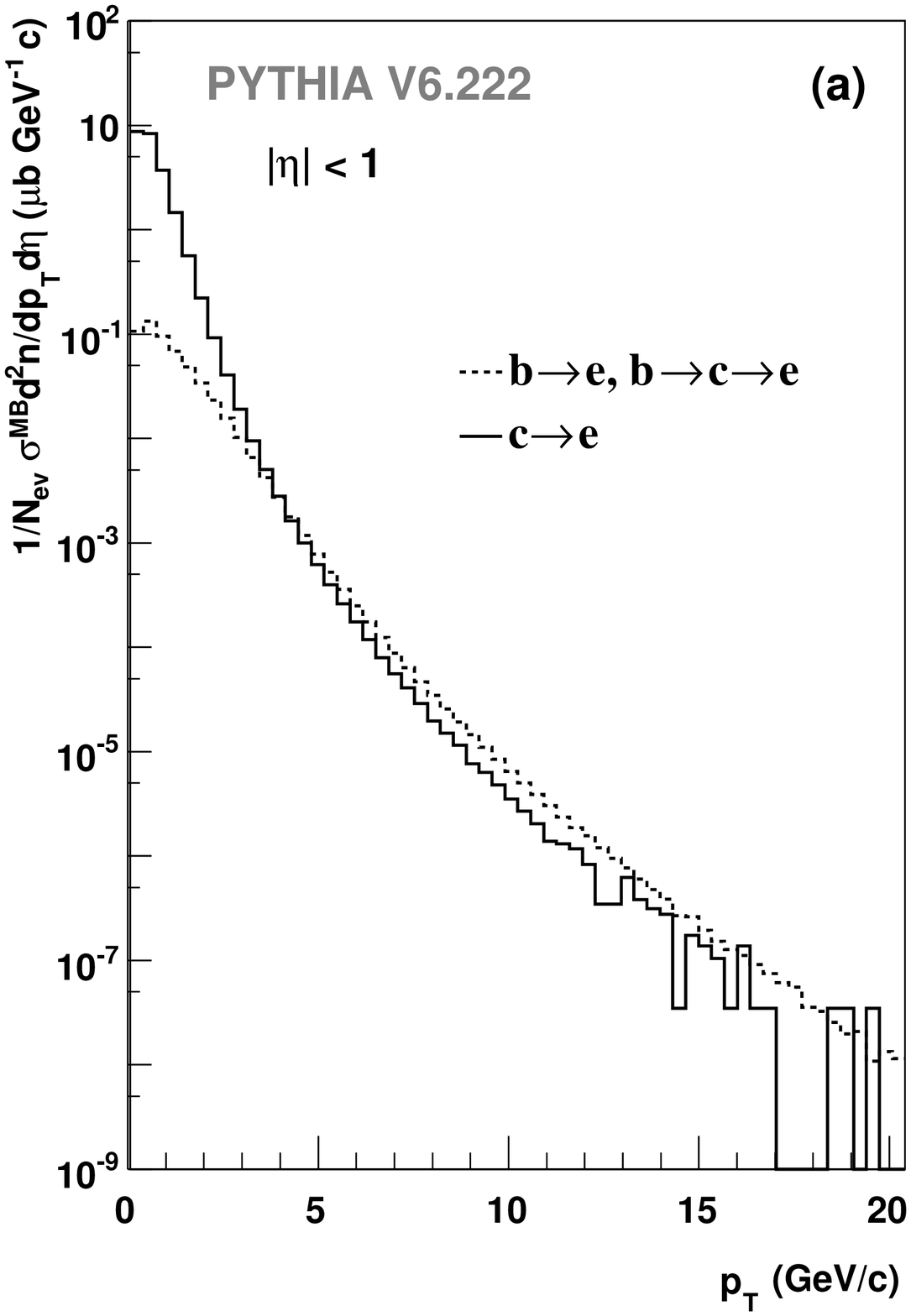}}
\subfigure{\includegraphics[width=0.23\textwidth]{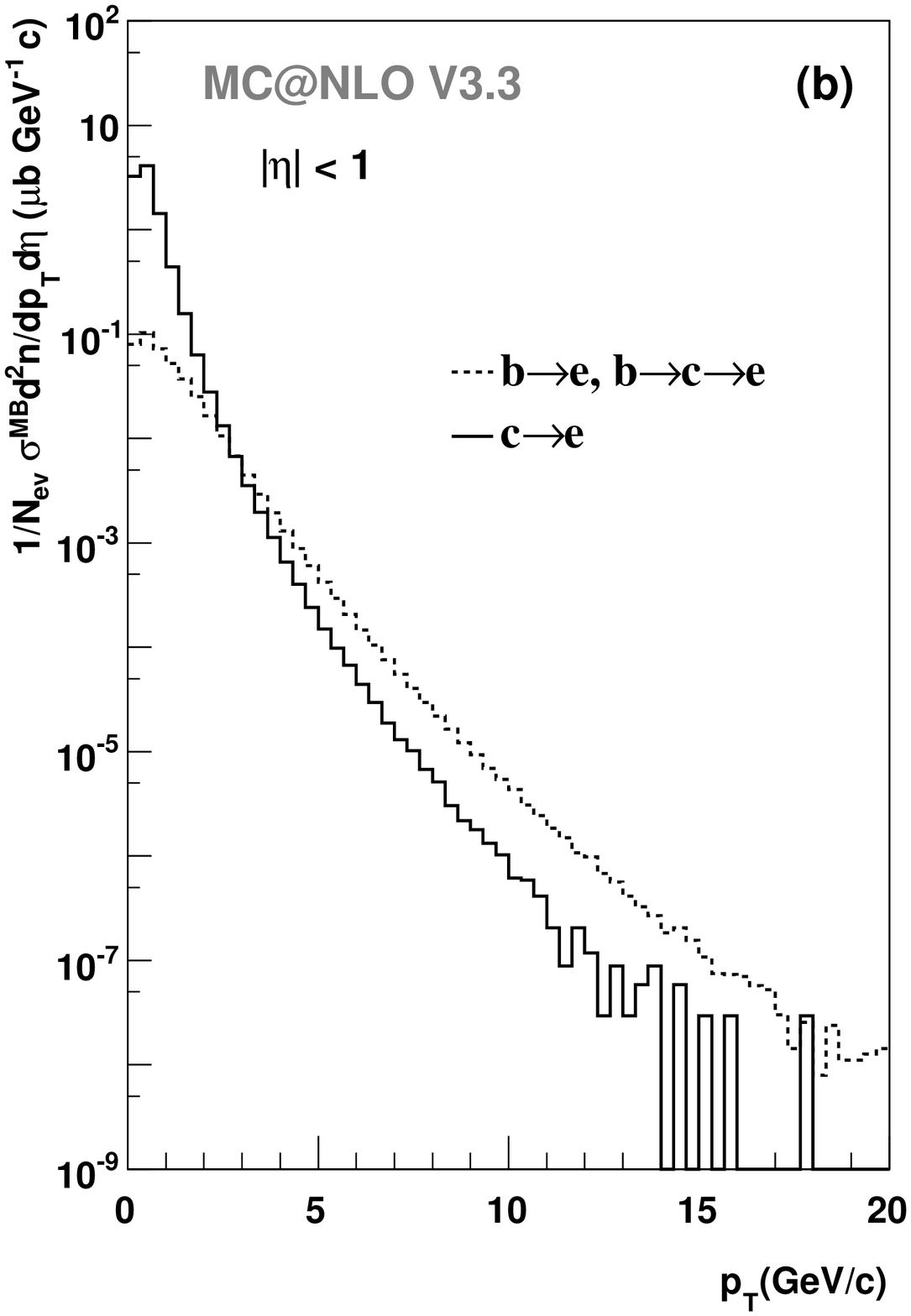}}
\caption{
Transverse momentum spectrum of charm (solid line) and bottom decay electrons (dashed line) in (a) PYTHIA and (b) MC@NLO simulations of $p+p$ collisions at $\sqrt{s}$ = 200 GeV.}
\label{fig:1}
\end{center}
\end{figure}

A charge-sign condition on the trigger electron and decay kaon provides a powerful tool to separate events with a $c{\bar c}$ or a $b{\bar b}$ pair.
As an example, Figs.~\ref{fig:11} (a) and (b) illustrate a schematic view of the fragmentation of a $c{\bar c}$ and a $b{\bar b}$ pair, respectively.
Assuming the trigger lepton is an electron from the fragmentation of a ${\bar c}$ or $b$ quark, the partner charm quark must be a $c$, hence producing a $K^-\pi^+$ pair. 
The bottom quark on the opposite side is a ${\bar b}$, which yield $K^+\pi^-$ pairs via the main decay mode $B \rightarrow \overline{D^0} + X$ (BR = 59.6$\%$). 
However, there is another channel, $B \rightarrow D^0 + X$ (BR = 9.1$\%$), which give $K^-\pi^+$ pairs~\cite{PDG}.
$e^-K^-$ ($e^+K^+$) pairs are also expected from semileptonic $B$ decays, e.g., $B^- \rightarrow D^0 e^- \overline{\nu_e}$.

Thus, electron$-$kaon pairs with the opposite charge sign (called unlike-sign $e-K$ pairs) identify B decays on the away-side of the azimuthal correlation distribution of decay electrons and $D^0$ mesons.
Requiring like-sign $e-K$ pairs select bottom on the near-side and charm and a small contribution from bottom  ($\sim$15$\%$) on the away-side of the $e-D^0$ correlation function.

Requiring $e-D^0$ coincidence in the same event significantly improves the signal-to-background ratio over either technique individually. 
Moreover, the decay electrons provide an efficient trigger for heavy-quark production events.
The shape of the azimuthal correlation distribution allows a more differential comparison between the charm and bottom contributions owing to their different decay kinematics.
The feasibility for this correlation method is examined using PYTHIA and MC@NLO simulations.

\begin{figure}[t]
\begin{center}
\subfigure{\includegraphics[width=0.23\textwidth]{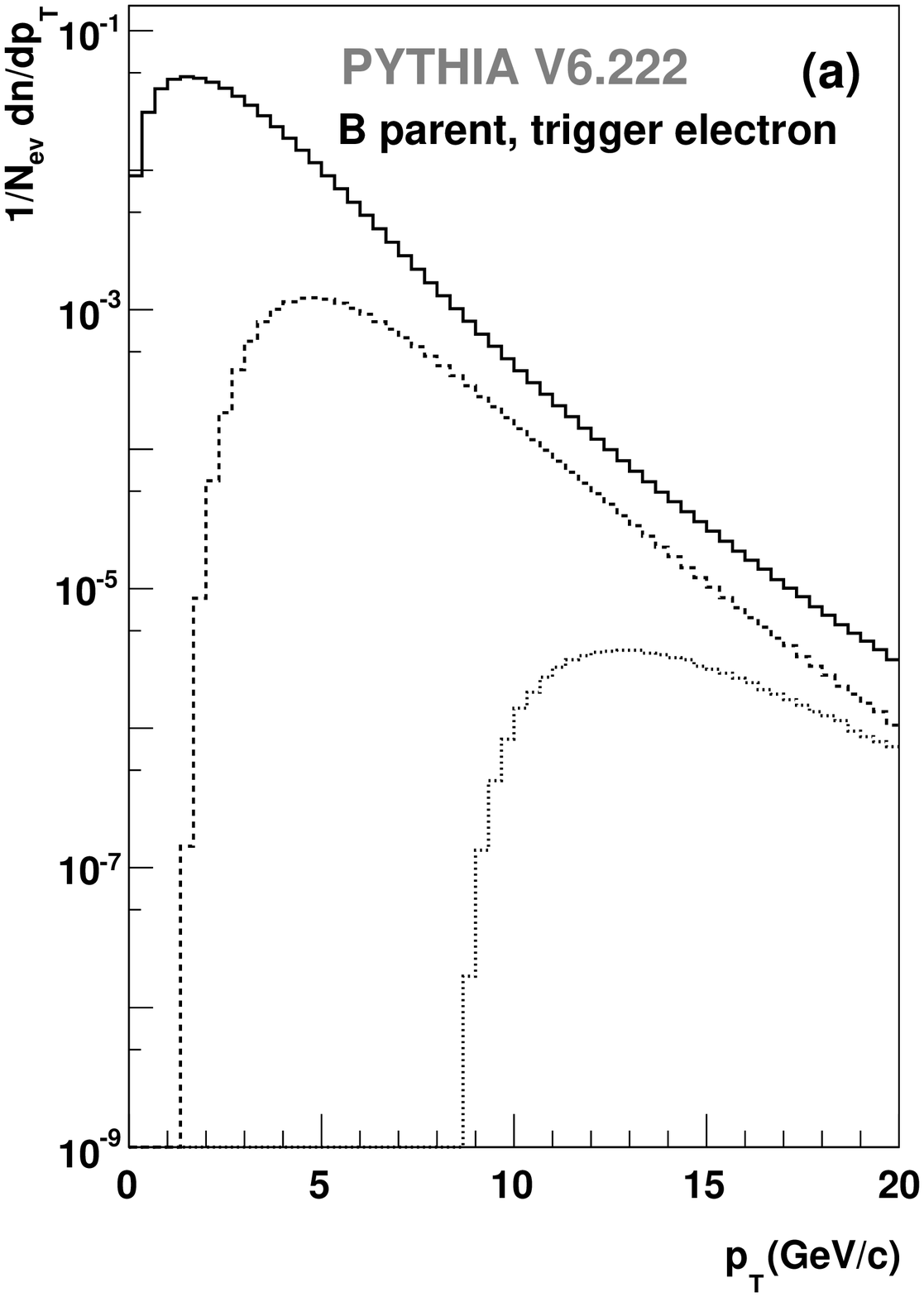}}
\subfigure{\includegraphics[width=0.23\textwidth]{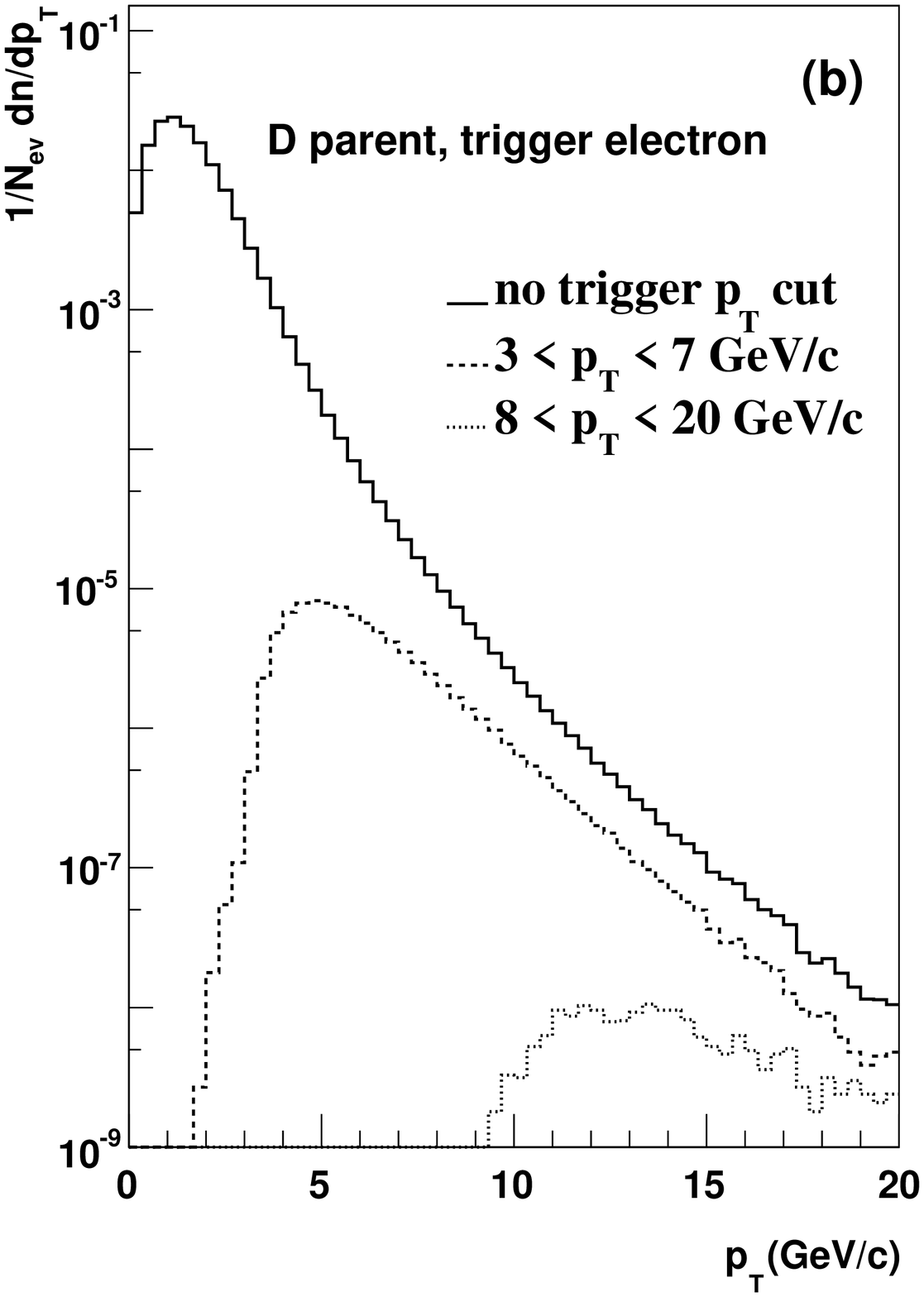}}
\subfigure{\includegraphics[width=0.23\textwidth]{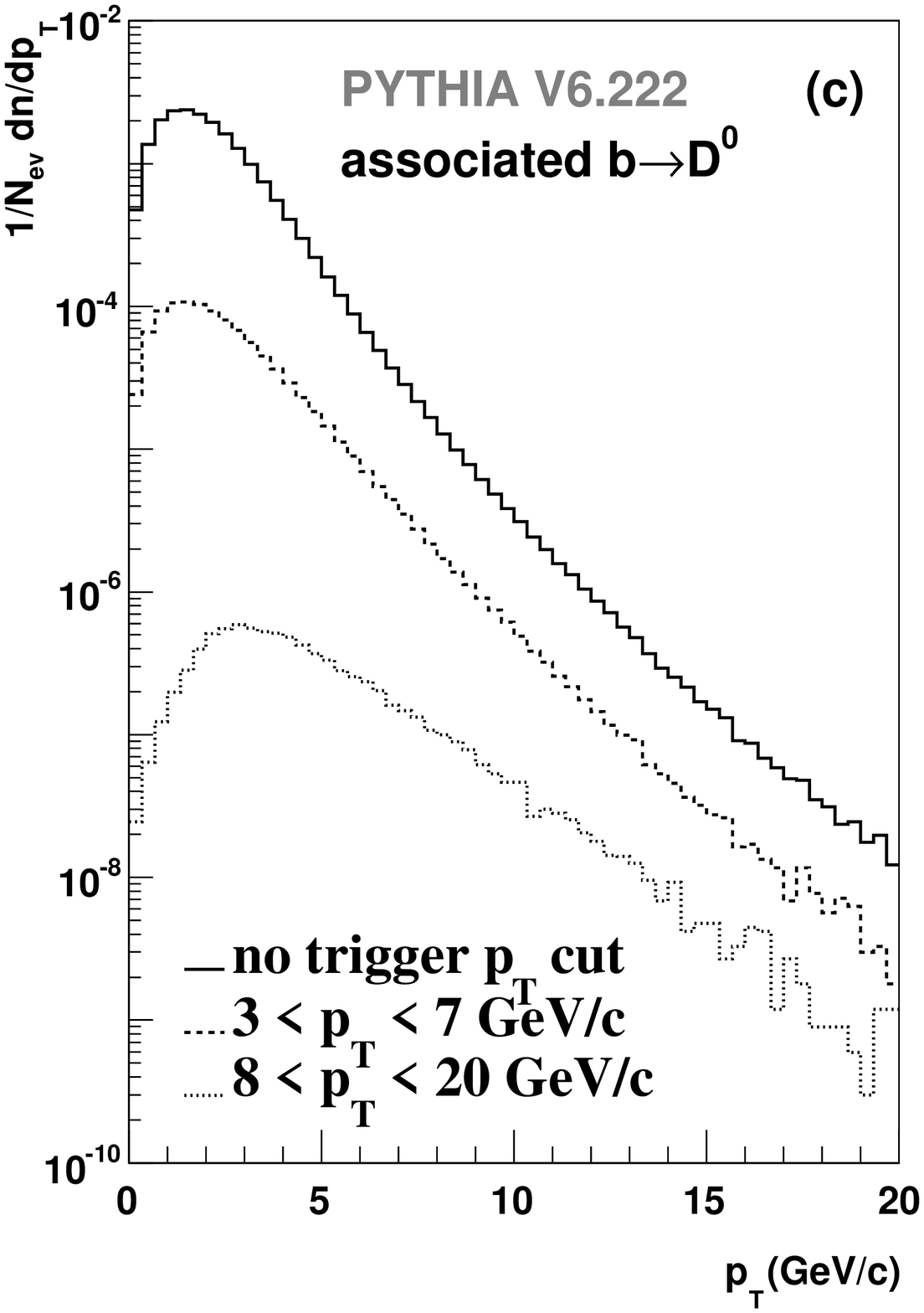}}
\subfigure{\includegraphics[width=0.23\textwidth]{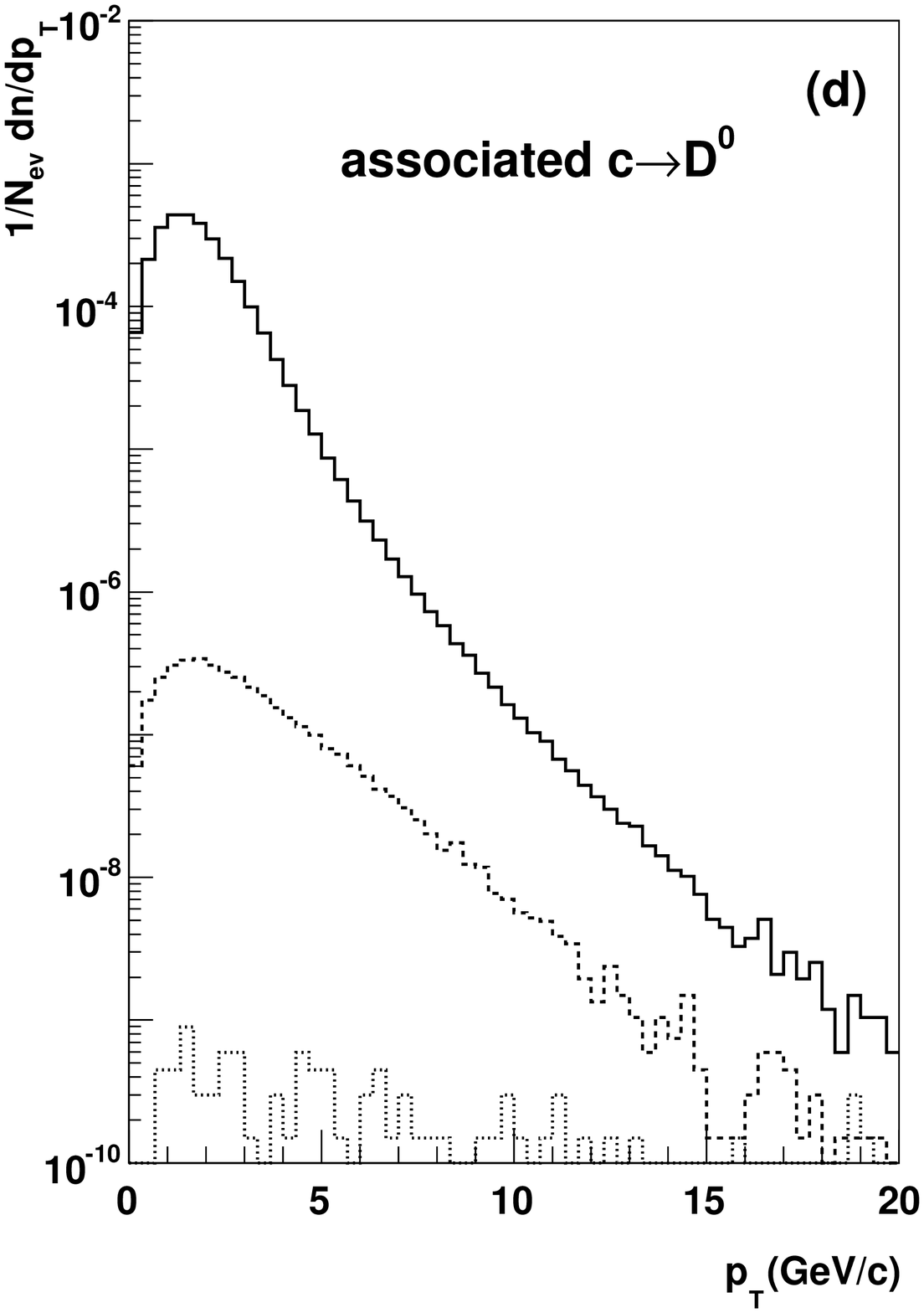}}
\caption{
Transverse momentum distribution of the (a) $B$ and (b) $D$ mesons that yield trigger electrons in the indicated $p_T$ ranges. 
Panels (c) and (d) illustrate the transverse momentum distribution of the $D^0$ mesons from bottom and charm fragmentation, respectively, opposite the trigger electrons in the specified $p_T$ ranges. The electrons and $D^0$ decay products, kaon and pion, are selected having a pseudo-rapidity $|\eta| < 1$.}
\label{fig:2}
\end{center}
\end{figure}

\section{Monte Carlo simulations}
The angular correlation function of charm and bottom decay electrons and $D^0$ mesons has been studied using leading-order PYTHIA simulations (version 6.222 with CTEQ5L PDF set, $m_c$ = 1.3 GeV/c$^2$ and $m_b$ = 4.5 GeV/c$^2$)~\cite{Mod:pythia} of $p+p$ collisions at $\sqrt{s}$~= 200 GeV.
In total, 8 billion events are generated for charm ($MSEL =$ 4) and 4 billion events for bottom production ($MSEL =$ 5) with a cross section of 232 and 2.13 $\mu$b, respectively. The default Peterson fragmentation function is used and the $D/D^*$ spin factor is taken into account~\cite{David}.
Electrons within the pseudo-rapidity range of $|\eta| < 1$ are assigned as trigger particles if they originate from charm ($D^{0}$, $D^{+}$, $D_{s}^{+}$ or their excited states) or bottom meson decays ($B^{0}$, $B^{+}$, $B_{s}^{0}$ or their excited states).
The transverse momentum ($\pT$) distribution of the decay electrons is shown in Fig.~\ref{fig:1}(a). 
Electrons from bottom decays starts to dominate over electrons from charm decays above $\pT \gtrsim$ 4 GeV/c, consistent with results from pQCD calculations at the fixed-order plus next-to-leading log (FONLL) level ~\cite{Theo:Matteo,Vogt08}.
Figures~\ref{fig:2}(a) and (b) depict the $\pT$ spectrum of $B$ and $D$ mesons, respectively, that yield trigger electrons in the indicated $\pT$ ranges.
The associated $D^0$ mesons are accepted if their decay products (kaon and pion) fall within the pseudo-rapidity window $|\eta| < 1$.
Figures~\ref{fig:2}(c) and (d) illustrate the $\pT$ distribution of the associated $D^0$ mesons from bottom and charm fragmentations, respectively.\\

\begin{figure}[t]
\begin{center}
\includegraphics[width=0.5\textwidth]{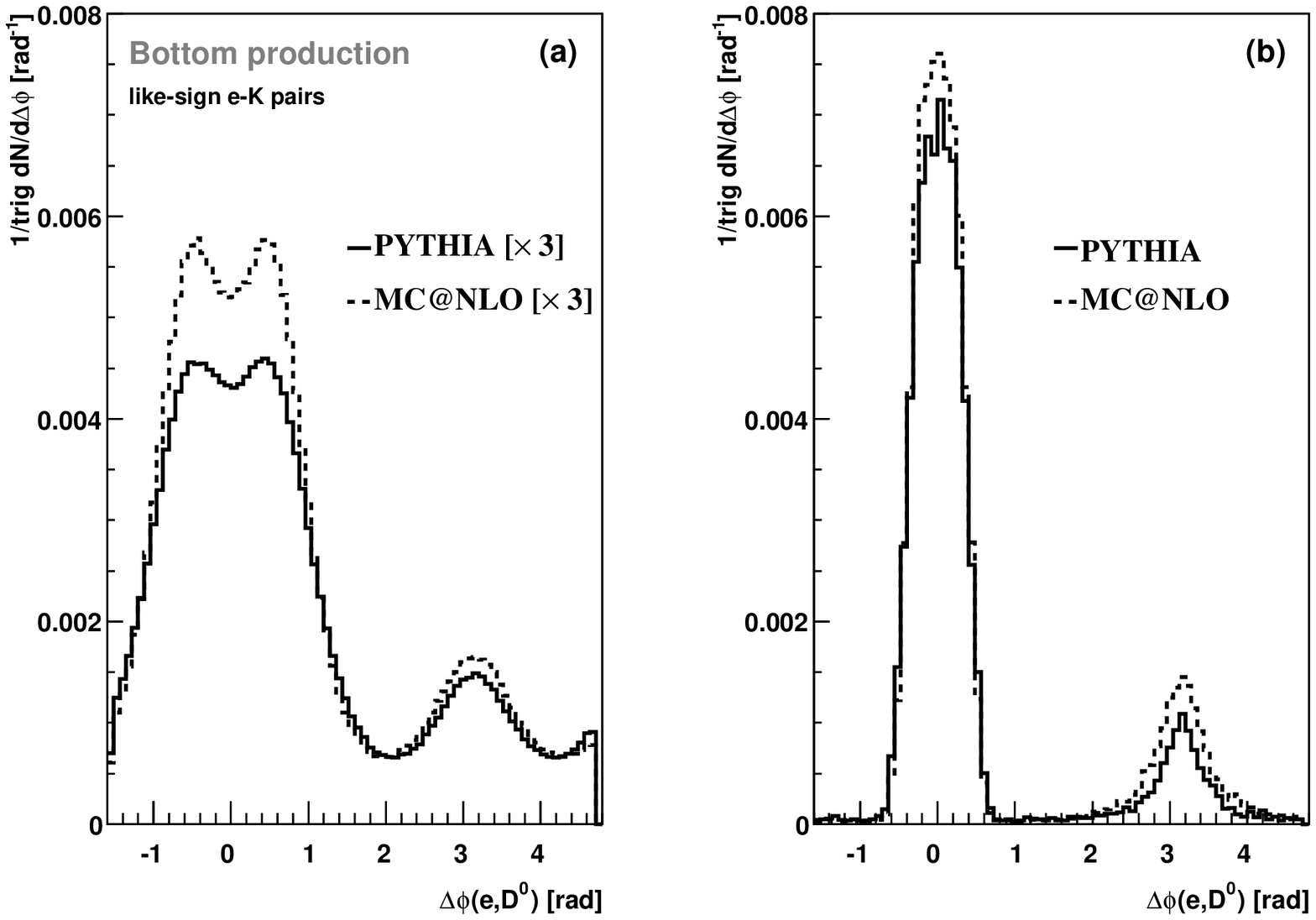}
\includegraphics[width=0.5\textwidth]{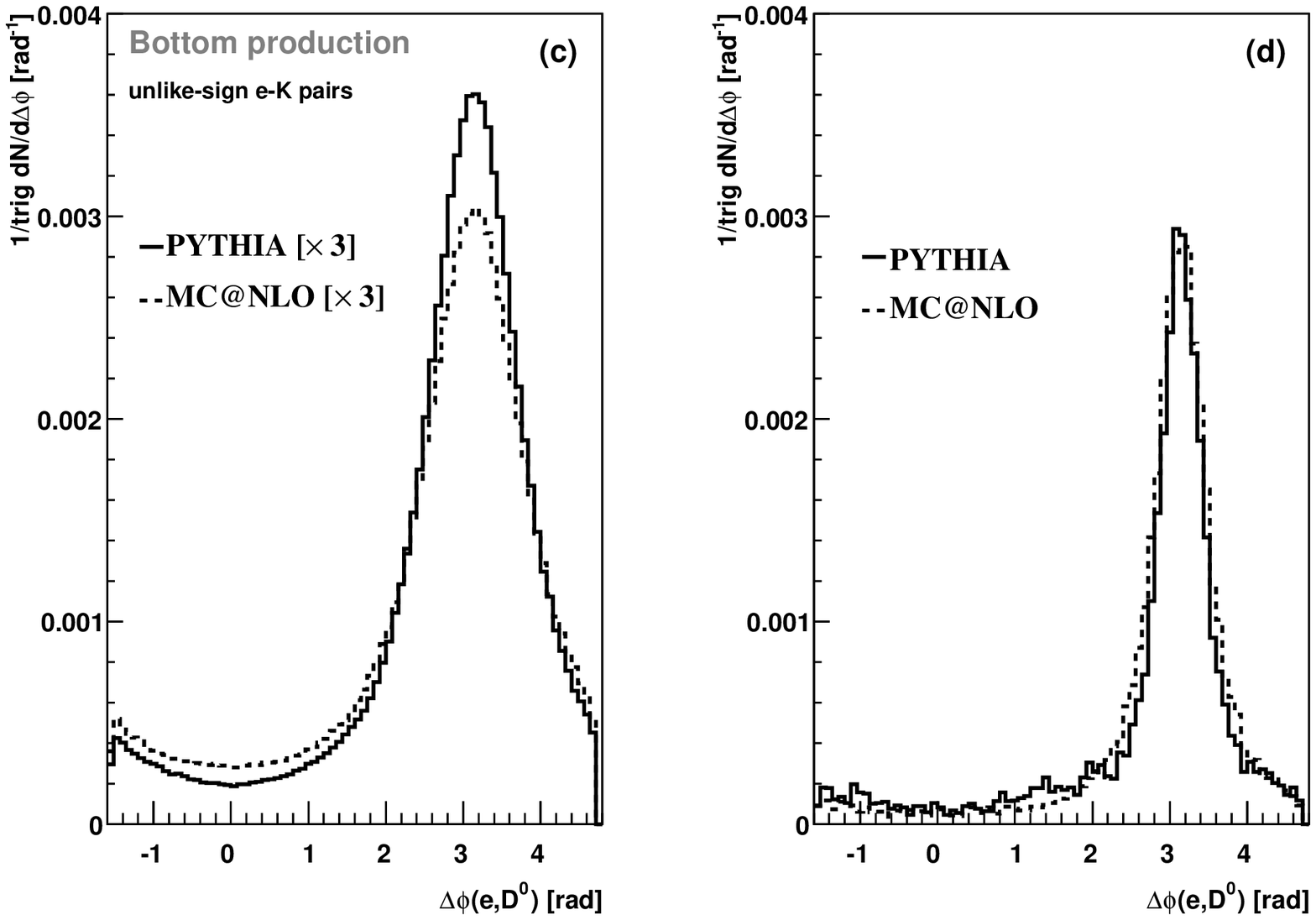}
\caption{
Azimuthal angular correlation distribution of electrons and $D^0$ mesons from bottom decays generated in PYTHIA (solid line) and MC@NLO simulations (dashed line) of 200 GeV $p+p$ collisions requiring like-sign (upper panels) and unlike-sign $e- K$ pairs (lower panels). 
The distributions are shown for trigger-electron transverse momentum ranges of (a+c) $3 < \pT < 7$ GeV/c and (b+d) $7 < \pT < 20$ GeV/c.}
\label{fig:3}
\end{center}
\end{figure}

\begin{figure}[t]
\begin{center}
\subfigure{\includegraphics[width=0.235\textwidth]{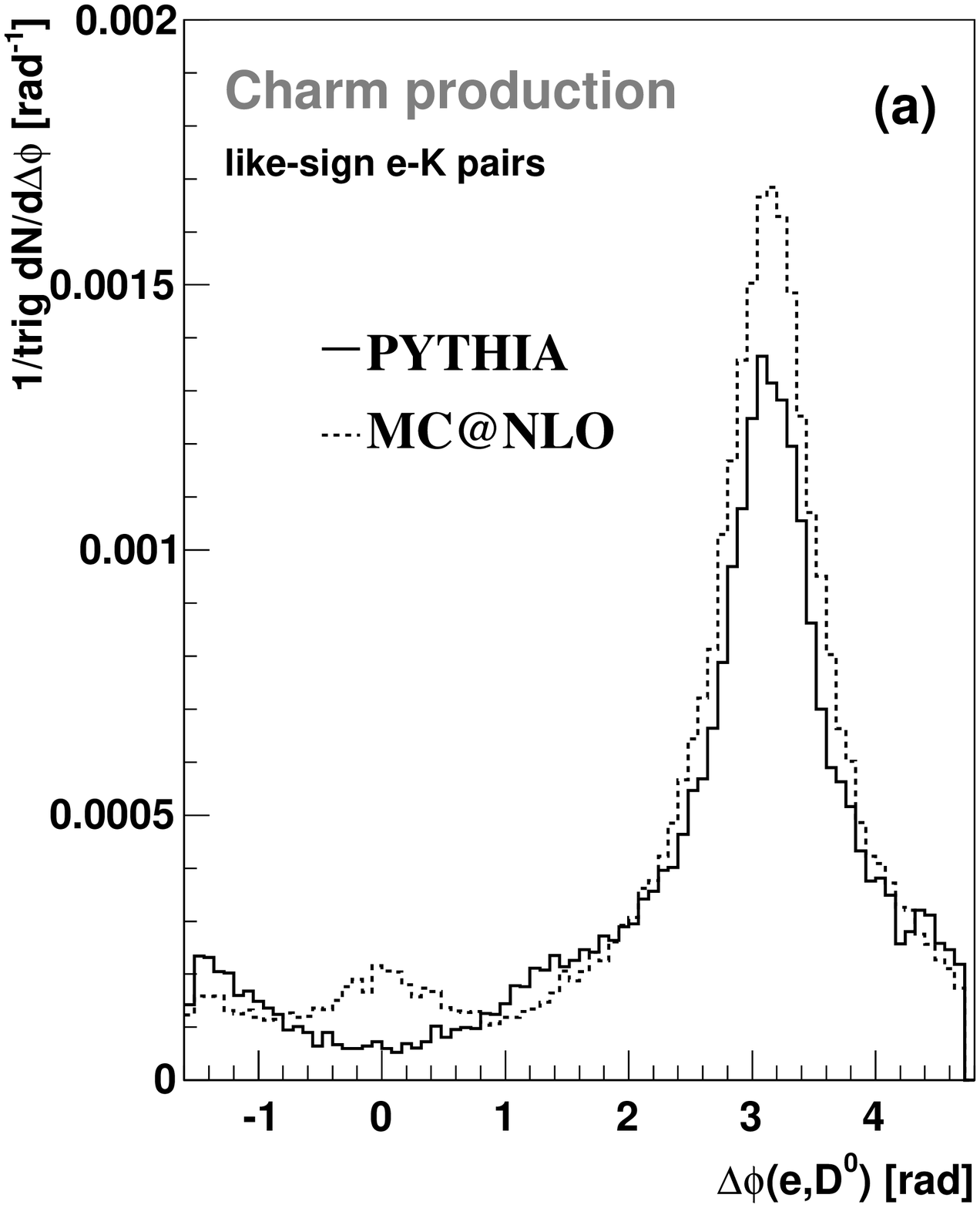}}
\subfigure{\includegraphics[width=0.235\textwidth]{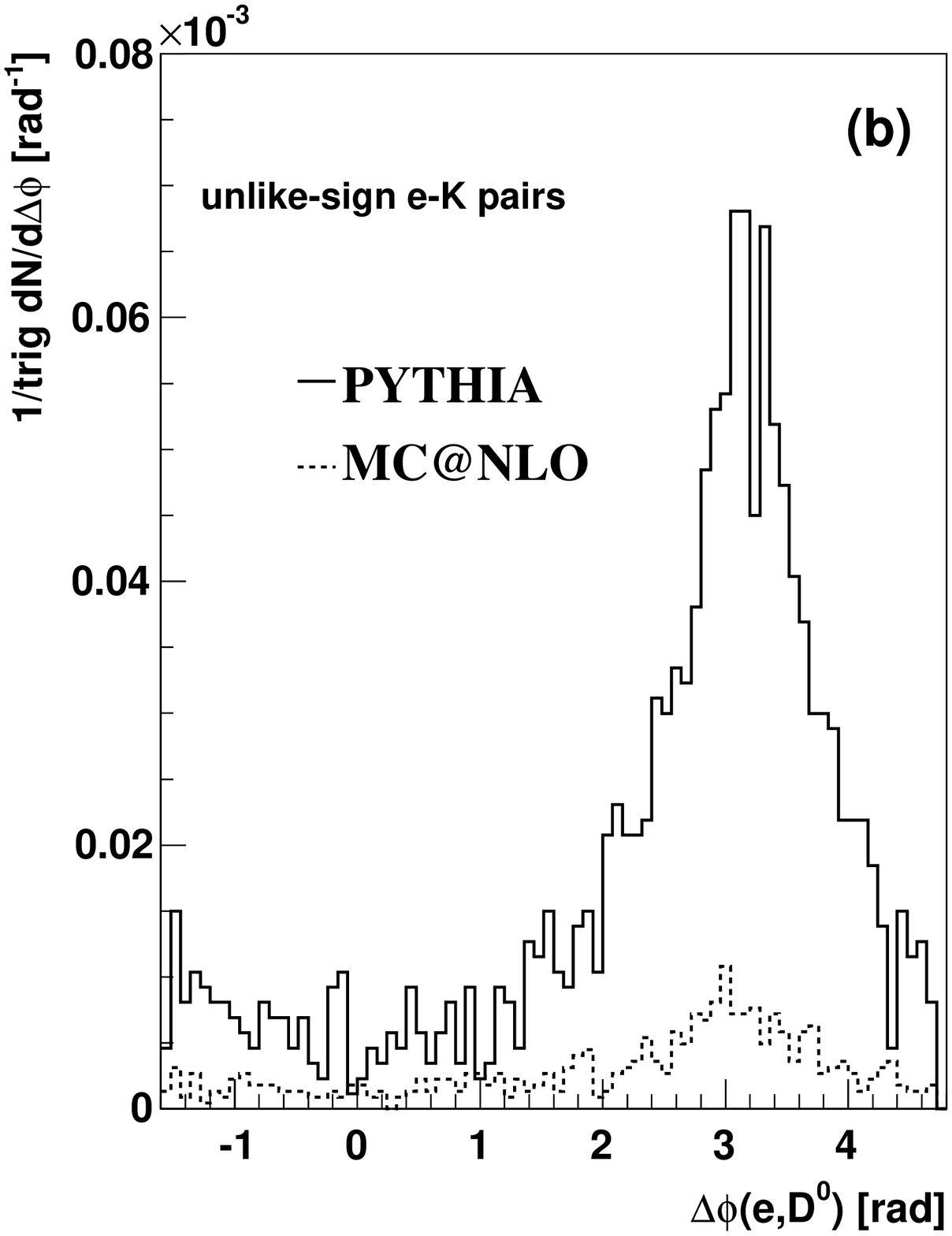}}
\caption{
Azimuthal angular correlation distribution of electrons and $D^0$ mesons from charm decays obtained from PYTHIA (solid line) and MC@NLO simulations (dashed line) for (a) like- and (b) unlike-sign $e- K$ pairs. 
Trigger-electron $\pT$ range is $3 < \pT < 7$ GeV/c.
Note the different scales for the correlation yield.}
\label{fig:4}
\end{center}
\end{figure}

The azimuthal correlation function is calculated for all electron$-D^0$ and positron$-\overline{D^0}$ pair combinations assuming a $D^0$ reconstruction efficiency of $\sim$70$\%$ as typically observed in large acceptance experiments like the STAR detector~\cite{Star:d0dAu}.
In the following, we imply electron$-D^0$ and positron$-\overline{D^0}$ pairs when using $e-D^0$.
Figures~\ref{fig:3}(a) and (b) show the azimuthal correlation distribution of heavy-quark decay electrons and $D^0$ mesons for like-sign $e-K$ pairs from bottom production for two different trigger-electron $\pT$ ranges.
The same $e-D^0$ correlation distribution is depicted in Fig.~\ref{fig:3}(c) and (d) for unlike-sign $e-K$ pairs from bottom production and in Figs.~\ref{fig:4}(a) and (b) for like and unlike-sign $e-K$ pairs  from charm production, respectively.
Comparing the upper and lower panels of Fig.~\ref{fig:3} one can conclude that like-sign $e-K$ pairs select $D^0$ mesons from $B$ decays on the near-side correlation whereas unlike-sign $e-K$ pairs separate $D^0$ mesons from $b\bar{b}$ flavor creation on the away-side correlation.
The near-side peak from $B$ decays is relatively broad at intermediate $\pT$ (3 $< \pT <$ 7 GeV/c) and exhibits a double peak structure (cf. Fig.~\ref{fig:3}(a)) which vanishes at higher $\pT$ (cf. Fig.~\ref{fig:3}(b)).
A comparison of the Figs.~\ref{fig:3}(a) and ~\ref{fig:4}(a) indicates that, for like-sign $e-K$ pairs, the near-side peak is dominated by $D^0$ mesons from $B$ decays whereas the away-side peak stems mainly from charm pair production (flavor creation).
The charm contribution for unlike-sign $e-K$ pairs on the away-side is small ($\sim$14$\%$ compared to the like-sign $e-K$ pairs) as shown in Figs.~\ref{fig:4}(b).\\

It has been shown~\cite{Field, Norrbin} that higher order sub-processes like gluon splitting may have a significant contribution to the near-side correlation. 
The contribution from gluon splitting was determined using MC@NLO simulations of $p+p$ collisions (version 3.3 with CTEQ6M PDF set) which allows modeling heavy-flavor hadro-production in a next-to-leading-order approach~\cite{Mod:mcnlo}. 
The MC@NLO computation uses the HERWIG event generator (version 6.510)~\cite{Mod:herwig} for parton showering, hadronization and particle decays.
1 billion events are generated for each charm and bottom production with a cross section of 184 and 1.6 $\mu$b, respectively.
The same particle selection criteria are used as for the PYTHIA simulations.
The $\pT$ spectrum of heavy-quark decay electrons is illustrated in Fig.~\ref{fig:1}(b).
Bottom decay electrons starts to dominate over charm decay electrons at a slightly lower $\pT$ compared to the PYTHIA results (cf. Fig.~\ref{fig:1}(a)). This seems to be due to the softer $\pT$ spectrum of the electrons from charm decays in the MC@NLO calculations.

Figures~\ref{fig:3} and~\ref{fig:4} also show the results from MC@NLO simulations for the trigger normalized angular correlation function of electrons and $D^0$ mesons from bottom and charm production events, respectively.
The correlation distribution from bottom production exhibits a similar shape as observed for PYTHIA simulations (cf. Figs.~\ref{fig:3}(a-d)).
The away-side peak shape of the correlation function from charm production (cf. Fig.~\ref{fig:4}(a)) agrees within 10-20$\%$ with the results from PYTHIA simulations.
This agreement is remarkable since these two event generators use different models for parton showering and hadronization ($k_{t}$ ordering in shower and string hadronization for PYTHIA and angular-ordered shower and cluster hadronization for HERWIG).
The difference of the near-side peak in Fig.~\ref{fig:4}(a) can be attributed to gluon splitting and is found to be (6.5$\pm$0.5)$\%$ of the open charm production observed in the studied $\pT$ range.

\begin{figure}[t]
\begin{center}
\includegraphics[width=0.46\textwidth]{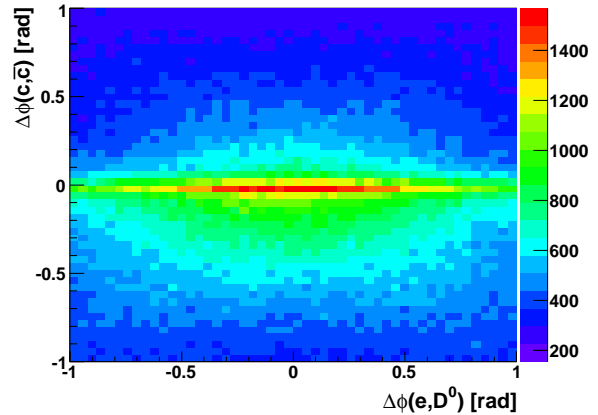}
\caption{
(Color online) Azimuthal correlation of $c\bar{c}$ pairs around the near-side azimuthal correlation of $e-D^0$ pairs obtained from MC@NLO simulations.}
\label{fig:5}
\end{center}
\end{figure}

Figure~\ref{fig:5} depicts a two-dimensional plot showing the azimuthal correlation distribution of $c{\bar c}$ pairs ($\dphi(c{\bar c})$) around the near-side peak of the azimuthal correlation distribution of $e-D^0$ pairs ($\dphi(e, D^0)$). The $\dphi(c{\bar c})$ distribution exhibits a clear peak around zero which supports the assumption that the near-side correlation peak of the $\dphi(e, D^0)$ distribution is indeed from gluon splitting.

\section{Extraction of the relative bottom contribution}
%
The relative bottom contribution for trigger electrons in the kinematical range $3 < \pT < 7$ GeV/c is obtained in two ways by comparison of the $e-D^0$ correlation yield on the near- ($\Delta\phi = 0\pm\pi/2$) and away-side ($\Delta\phi =\pi\pm\pi/2$) from Figs.~\ref{fig:3}(a+c) and~\ref{fig:4}(a). 

Firstly, by requiring like-sign $e-K$ pairs which selects bottom on the near-side (cf. Fig.~\ref{fig:3}(a)) and charm on the away-side (cf. Fig.~\ref{fig:4}(a)). 
The relative bottom contribution $\frac{e_{B}}{e_{B}+e_{D}}$ is obtained from the $D^0$ yield on the near-side in Fig.~\ref{fig:3}(a) ($D^0({\rm NS, b)}$) and away-side in Fig.~\ref{fig:4}(a) ($D^0({\rm AS, c)}$) according to
\[ \frac{e_{B}}{e_{B}+e_{D}} = \frac{1}{1+\frac{\frac{D^0({\rm AS, c})}{BR (c\rightarrow D^0+X)}}{D^0({\rm NS, b})}}. \]
The branching ratio $BR$ takes into account that $D^0$ from semileptonic bottom decays are always accompanied by an electron or more general by a lepton whereas electrons from charm decays have a probability of 56.5$\%$ to be balanced by a $D^0$ meson.
The $\frac{e_{B}}{e_{B}+e_{D}}$ ratio is found to be 0.52$\pm$0.03 for PYTHIA and MC@NLO simulations.

Secondly, the relative bottom contribution is determined from the $D^0$ yield on the away-side which selects charm for like-sign $e-K$ pairs (cf. Fig.~\ref{fig:4}(a)) and bottom for unlike-sign $e-K$ pairs (cf. Fig.~\ref{fig:3}(c)). 
The $c/b$ ratio is determined from the away-side $D^0$ correlation yield in Fig.~\ref{fig:4}(a) ($D^0$(LS, c)) and Fig.~\ref{fig:3}(c) ($D^0$(ULS, b)) by
\[ \frac{D^0({\rm LS, c})}{D^0({\rm ULS, b})} = c/b \times \frac{BR (c\rightarrow D^0+X)}{BR (b\rightarrow D^0+X)}. \]
PYTHIA and MC@NLO simulations give a $c/b$ ratio of 1.01$\pm$0.07 and 1.27$\pm$0.09, respectively, for trigger electrons in the $\pT$ range $3 < \pT < 7$ GeV/c.
From
\[ \frac{e_{B}}{e_{B}+e_{D}} = \frac{1}{1+c/b \times \frac{BR (c\rightarrow e+X)}{BR (b\rightarrow e+X)}}, \]
\noindent 
where the branching ratios for the $c$ and $b$ decays to electrons are quite similar,
the $\frac{e_{B}}{e_{B}+e_{D}}$ is found to be 0.53$\pm$0.05 and 0.47$\pm$0.04 for PYTHIA and MC@NLO, respectively. The uncertainties are obtained from the sum of the experimental uncertainties of the branching fractions in quadrature.

The results obtained with the two different approaches agree within uncertainties. Furthermore, the extracted $\frac{e_{B}}{e_{B}+e_{D}}$ ratios show agreement with the relative bottom contribution from FONLL calculations~\cite{Theo:Matteo,Vogt08}.

\section{Summary}
The azimuthal angular correlation of heavy-flavor decay electrons and $D^0$ mesons in combination with a charge-sign requirement on electron and $D^0$-decay kaon pairs allows, on a statistical basis, the separation of charm and bottom production and their sub-processes.
The feasibility for this new correlation method is shown using PYTHIA and MC@NLO simulations which also yield an estimate of the complete NLO contributions (including gluon-splitting diagrams).
The relative bottom contribution to the heavy-flavor decay electrons is determined by comparison of the near- and away-side correlation distributions for charm and bottom production processes and is found to be $\sim$50 $\%$ in the studied transverse momentum range $3 < \pT < 7$ GeV/c .

\begin{acknowledgments}
Acknowledgments: 
The author thanks M. Cacciari, S. Frixione, T. Ullrich, R. Vogt and R. Kamermans for fruitful discussions. 
The European Research Council has provided financial support under the European Community's Seventh Framework Programme (FP7/2007-2013)/ ERC grant agreement no 210223.
This work is supported in part by a Veni grant from the Netherlands Organization for Scientific Research (project number: 680-47-109).
\end{acknowledgments}


\end{document}